\documentclass[10pt,a4paper,superscriptaddress,aps,prl,twocolumn,floatfix,citeautoscript,showpacs]{revtex4-1}
\usepackage{amsfonts,amssymb,amsmath,latexsym}
\usepackage{graphicx}
\usepackage{xcolor}
\usepackage{setspace}
\usepackage{setspace}
\usepackage{graphicx}
\usepackage{multirow}
\usepackage{tabularx}
\usepackage{natbib}
\usepackage{mciteplus}
 \usepackage[margin=0.80in]{geometry}
\newcommand{\br}{\mathbf{r}}

\newcommand{\imun}{\mathrm{i}}

\newcommand{\lcpq}{Laboratoire de Chimie et Physique Quantiques, IRSAMC, CNRS, Universit\'e de Toulouse, UPS, France}
\newcommand{\etsf}{European Theoretical Spectroscopy Facility (ETSF)}
\newcommand{\unibo}{Universit\`a di Bologna, Bologna, Italy}
\newcommand{\perugia}{Dipartimento di Chimica, Biologia e Biotecnologie, Universit\`a degli Studi di Perugia, Via Elce di Sotto 8, 06123 Perugia, Italy}

\begin{document}
\title{ 
A simple position operator for periodic systems
}

\author{Emilia Aragao Valen\c{c}a}
\author{Diego Moreno}
\affiliation{\lcpq}
\author{Stefano Battaglia}
\affiliation{\lcpq}
\affiliation{\perugia}
\author{Gian Luigi Bendazzoli}
\affiliation{\unibo}
\author{Stefano Evangelisti}
\email{stefano.evangelisti@irsamc.ups-tlse.fr}
\author{Thierry Leininger}
\affiliation{\lcpq}
\author{Nicolas Suaud}
\affiliation{\lcpq}
\author{J.~A.~Berger}
\email{arjan.berger@irsamc.ups-tlse.fr}
\affiliation{\lcpq}
\affiliation{\etsf}

\date{\today}

\begin{abstract}
We present a position operator that is compatible with periodic boundary conditions (PBC).
It is a one-body operator that can be applied in calculations of correlated materials by simply replacing the traditional position vector 
by the new definition. We show that it satisfies important fundamental as well as practical constraints.
To illustrate the usefulness of the PBC position operator we apply it to the localization tensor, a key quantity that
is able to differentiate metallic from insulating states.
In particular, we show that the localization tensor given in terms of the PBC position operator 
yields the correct expression in the thermodynamic limit.
Moreover, we show that it correctly distinguishes between finite precursors of metals and insulators.
\end{abstract}

\maketitle

\renewcommand{\baselinestretch}{1.5}

Expectation values that involve the position operator $\hat{\br}$ play a prominent role in both molecular and condensed-matter physics.
Many important quantities are expressed in term of $\hat{\br}$, \emph{e.g.}, the multipole moments and the localization tensor.
The latter quantity was introduced by Resta and co-workers~\cite{Resta_1999,Sgiarovello_2001,RestaJCP} 
following an idea of Kohn~\cite{Kohn} that information about electron localization should be obtained from the ground-state wave function
(see also Ref.~\cite{Kudinov}).
The localization tensor is able to distinguish between conductors and insulators:
when the number of electrons tends to infinity it diverges in the case of a conductor, while it remains finite in the case of an insulator.
It has been applied to study the metallic behavior of clusters~\cite{vetere_2008,Bendazzoli2011,giner,monari2008,Evangelisti2010,Bendazzoli2010,Bendazzoli2012,ElKhatib2015,fertitta2015}
and has recently also been used to investigate Wigner localization~\cite{Diaz-Marquez_2018}.

In its standard definition the position operator $\hat\br$ is simply defined as the multiplication with the position vector $\br$.
However, this definition is not compatible with periodic boundary conditions (PBC), since $\br$ is not a periodic function.
This is a problem, since many quantities of interest are related to the solid state which are conveniently described using PBC.
Therefore, it is of great interest to search for a position operator that is compatible with PBC while reducing to the position vector $\br$ in the appropriate limit.
We will provide such a definition of the position operator in this work.
For notational convenience we will consider only one dimension in the remainder of this work.
All our findings can be generalized to higher dimensions.
We use Hartree atomic units ($\hbar=1$, $e=1$, $m_e=1$, $4\pi\epsilon_0=1$).

We study a system of length $L$ whose electronic many-body wave function $\Psi(t)$ satisfies PBC, \emph{i.e.}, for each $x_i$
the following condition holds,
\begin{equation}
\Psi(x_1,\cdots\!,x_i,\cdots\!,x_N)\!=\!\Psi(x_1,\cdots\!,x_i + L,\cdots\!,x_N),
\end{equation}
where $N$ is the number of electrons.
We are looking for a position operator that is compatible with PBC.
We denote such an operator as $\hat{q}$.

Let us summarize important criteria that $\hat{q}$ should satisfy:
1) $\hat{q}$ should be invariant with respect to a translation $L$;
2) $\hat{q}$ should reduce to the standard position operator $\hat{x}=x$
for finite systems described within PBC, 
\emph{i.e.}, in a supercell approach ($L\rightarrow\infty$ for fixed $N$)~\cite{Payne_1992}
one should obtain results that coincide with those obtained within open-boundary conditions (OBC).
3) The distance defined in terms of $\hat{q}$ should be gauge-invariant, \emph{i.e.}, it should be independent of the choice of the origin.
This criterium is important since the main purpose of a position operator
is to yield the correct distance between two spatial coordinates.
Finally, we add a fourth criterium:
4) For a system of many particles, $\hat{q}$ should be a one-body operator, as is $\hat{x}$.
Although the last criterium is not a fundamental one, it is crucial if we want to apply 
the new operator to realistic systems.


In a seminal work Resta proposed a definition for the \emph{expectation value} 
of the total position operator $\hat{X} = \sum_{i=1}^N x_i$ that is compatible with PBC~\cite{Resta_1998}.
Following a similar strategy Resta and coworkers also proposed an expression for the localization tensor that is compatible with PBC~\cite{Resta_1999,Sgiarovello_2001}.
Despite the important progress made in these works there are also several shortcomings to this approach:
1) they provide definitions for expectation values \emph{but not a definition for the position operator itself};
2) the operators are $N$-body which make them unpractical for the calculation of expectation values of real correlated systems with many electrons.
Finally, we note that their localisation tensor is "formally infinite (even at finite $N$) in the metallic case"~\cite{Resta_1999}.
Therefore, their approach is not applicable to finite precursors of metals.

Instead, in this work we propose a definition for the \emph{position operator itself}.
We will demonstrate that it satisfies the four criteria mentioned above.
Moreover, we will explicitly show that it correctly yields the macroscopic polarization in the thermodynamic limit 
as well as a useful expression for the localization tensor.
The latter gives finite values at finite $N$ and $L$ while yielding the correct values in the thermodynamic limit.

In order to treat PBC systems, we associate to the electron position $x$ 
the {\em complex position} $q_L(x)$, defined as
\begin{equation}
q_L(x) = \frac{L}{2\pi \imun} \left[\exp\left({\frac{2\pi \imun}{L}x}\right) -1\right],
\label{Eqn:qdef}
\end{equation}
with $\imun$ the imaginary unit.
The complex position $q_L(x)$ is a continuous and infinitely derivable function of $x$.
In complete analogy with the quantum treatment of the position operator $\hat{x}$, 
we define the action of the complex position operator $\hat{q}$ as the multiplication with $q_L(x)$.

Let us review the criteria mentioned above with respect to $q_L(x)$.
1) $q_L(x)$ is trivially invariant under a translation, \emph{i.e.}, $q_L(x+L) = q_L(x)$.
Therefore the complex position $q_L(x)$, unlike the ordinary position $x$, satisfies the PBC constraint.
2) $q_L(x)$ reduces to the standard position operator $x$ in the limit $L\rightarrow\infty$, 
in the sense of a supercell approach mentioned above.
This can be shown by expanding the exponential function: $\exp\left({\frac{2\pi \imun}{L}x}\right) = 1+\frac{2\pi \imun}{L}x + O(1/L^2)$
\footnote{The supercell approach implies that whenever $x \approx L$ the wave function vanishes and, 
therefore, there is no contribution to expectation values.
Therefore, we can take the small parameter to be $1/L$}.
3) the distance defined in terms of $q_L(x)$ is gauge independent.
By defining the difference $q_{L,x_0}(x_2,x_1) = q_L(x_2-x_0) - q_L(x_1-x_0)$, where $x_0$ is the (arbitrary) origin,
it can be verified that
\begin{equation}
q_{L,x_0}(x_2,x_1) =
\frac{L}{2\pi \imun} e^{-{\frac{2\pi \imun}{L}x_0}} 
\left[e^{{\frac{2\pi \imun}{L}x_2}}-
e^{{\frac{2\pi \imun }{L}x_1}}
\right].
\end{equation}
Therefore, the distance $\sqrt{q^{\dagger}_{L,x_0}(x_2,x_1)q_{L,x_0}(x_2,x_1)}$ is independent of $x_0$, as it should.
We note that a proposition made in a short comment by Zak~\cite{Zak_2000}, in which the position operator is defined in terms of a sine function,
does not satisfy this important constraint and, therefore, the corresponding distance depends on the choice of the origin.
4) $q_L(x)$ is a one-body operator.

Let us now demonstrate how the complex position in Eq.~(\ref{Eqn:qdef}) yields a useful expression for the localization tensor within PBC
using an approach that is completely analogous to the OBC case.
The localization tensor $\lambda$ is defined as the total position spread (TPS) per electron where 
the TPS is a one-body operator that is defined
as the second cumulant moment of the total position operator $\hat{X} = \sum_{i=1}^N x_i$
\footnote{Even though in one dimension $\lambda$ is a scalar we use its general name, \emph{i.e.}, localization tensor, for consistency}:
\begin{equation}
\lambda(N) = \frac{1}{N}\left[\langle\Psi|\hat{X}^2 |\Psi\rangle  - \langle\Psi|\hat{X} |\Psi\rangle^2\right].
\label{Eqn:TPS_OBC}
\end{equation}
The second term in the square brackets ensures gauge invariance with respect to the choice of the origin of the coordinate system.
The localization tensor is translationally invariant.



In a complete analogy with the OBC definition, within PBC we replace the position of a particle $x_i$ by its complex position $q_L(x_i)$. 
In such a way, the complex total position operator is still a one-body operator, defined as
\begin{equation}
\hat{Q}_L = \sum_{i=1}^N \hat{q}_L(x_i).
\end{equation}

The localisation tensor within PBC $\lambda_L$ is also a real quantity, like $\lambda$.
It is defined as the second cumulant moment of the complex total position operator per electron:
\begin{equation}
\lambda_L(N) = \frac{1}{N}\left[\langle\Psi|\hat{Q}_L^{\dagger} \hat{Q}_L |\Psi\rangle - 
\langle\Psi|\hat{Q}_L^{\dagger}|\Psi\rangle \langle\Psi|\hat{Q}_L |\Psi\rangle\right]
\label{Eqn:TPS_PBC}
\end{equation}
in complete analogy with the OBC definition in Eq.~(\ref{Eqn:TPS_OBC}).
Equation ~(\ref{Eqn:TPS_PBC}), together with Eq.~(\ref{Eqn:qdef}), is the main result of this work.
The simple expression is completely general, it can be applied to correlated many-body wave functions and is valid for both finite systems and infinite systems, 
\emph{e.g.}, by taking the thermodynamic limit.
In the special case of a single-determinant wave function, it can be shown that the expression in Eq.~(\ref{Eqn:TPS_PBC}) coincides 
with the result obtained in Ref.~[\onlinecite{Sgiarovello_2001}] in the thermodynamic limit.


We will now demonstrate that the PBC localisation tensor given by Eq.~(\ref{Eqn:TPS_PBC}) can differentiate 
between metals and insulators and that it can be applied to both finite and infinite systems.
To do so we will apply it to model systems.
First, we will treat the simple case of $N$ non-interacting electrons in a one-dimensional box of length $L$.
This model system can be seen as a prototype of a conductor.
For this reason, it is particularly important that the formalism we propose in this work can be applied to such a system.
In analogy to the OBC case~\cite{RestaJCP}, one expects that the localization tensor $\lambda_L(N)$, diverges in the thermodynamic limit.

We consider $N  = 2m+1$ non-interacting electrons where $m$ is a non-negative integer. 
For the sake of simplicity, we assume that the particles are spinless.
In the case of particles with spin, the final result can be trivially obtained by multiplying the spinless-particle result
by the spin multiplicity of a single particle.
The eigenfunctions of the Hamiltonian of this system are periodic orbitals given by 
\begin{equation}
\phi_n(x) = \frac{1}{\sqrt[]{L}} \exp \left(\imun \frac{2\pi n}{L} x \right),
\label{Eqn:PiB_orbitals}
\end{equation}
where $n$ is an integer.
Since the particles do not interact, the ground-state wave function is a single Slater determinant of the occupied orbitals, 
given by $|\Phi_0\rangle = |\phi_{-m} \cdots \phi_{m}\rangle$.
The corresponding localisation tensor reads
\begin{align}
\notag
&\lambda_L(N) = \frac{L^2}{4\pi^2 N} \bigg[ \langle\Phi_0| \sum_{i=1}^N e^{-\frac{2\pi\imun}{L}x_i}
\sum_{j=1}^N e^{\frac{2\pi\imun}{L}x_j} |\Phi_0\rangle
\\ &-
\langle\Phi_0| \sum_{i=1}^N  e^{-\frac{2\pi\imun}{L}x_i} |\Phi_0\rangle
\langle\Phi_0| \sum_{j=1}^N e^{\frac{2\pi\imun}{L}x_j} |\Phi_0\rangle
\bigg].
\end{align}
Inserting a complete set of states in the first term in the square brackets yields
\begin{align}
&\lambda_L(N) \!=\!\! \frac{L^2}{4\pi^2 N}\!
\sum_{I\neq 0} \sum_{i,j=1}^N \!\!
\langle\Phi_0| e^{-\imun\frac{2\pi}{L}x_i} |\Phi_I\rangle
\langle\Phi_I| e^{\imun\frac{2\pi}{L}x_j} |\Phi_0\rangle
\\ & =
\frac{L^2}{4\pi^2 N}
\sum_{|p|\le m} \sum_{|l|> m} \langle\phi_p| e^{-\imun\frac{2\pi}{L}x}|\phi_l\rangle
\langle\phi_l|  e^{\imun\frac{2\pi}{L}x} |\phi_p\rangle,
\label{Eqn:SC}
\end{align}
where in the last step we used the Slater-Condon rules for one-electron operators~\cite{Szabo}.
Inserting Eq.~(\ref{Eqn:PiB_orbitals}) into the above expression leads to the following expression
for the matrix element $\langle\phi_p| e^{-\imun\frac{2\pi}{L}x}|\phi_l\rangle$,
\begin{equation}
\frac{1}{L}\int_L \exp\left(\imun\frac{2\pi(l-p-1)}{L}x\right) dx = \delta_{l-p-1}.
\end{equation}
Therefore, there is only one nonzero contribution in the double summation over $p$ and $l$ in Eq.~(\ref{Eqn:SC}),
namely when $p=m$ and $l=m+1$.
We can write the final result as
\begin{equation}
\lambda_L(N) =
\frac{L^2}{N^2}\frac{N}{4\pi^2}
\label{Eqn:LT},
\end{equation}
from which we can deduce the behavior of the localisation tensor in the thermodynamic limit.
Since, in this limit, $N/L$ remains constant, the localisation tensor diverges linearly with $N$, 
as one would expect for a measure of conductivity applied to a perfect conductor.


%

Finally, we consider a dimerized chain containing $4n+2$ atoms at half filling, \emph{i.e.}, $N=4n+2$, in a tight-binding model.
The Hamiltonian is given by
\begin{equation}
\hat{H} = \sum_{i=1}^N -t_i(a^{\dagger}_i a_{i+1} + a^{\dagger}_{i+1} a_i)
\label{Eqn:HTB}
\end{equation}
where $a_i^{\dagger}$ ($a_i$) is a creation (annihilation) operator and the hopping parameter $t_i = 1 - (-1)^i \delta$ with
$0\leq\delta\leq 1$. This means that the dimerization is at its maximum when $\delta=1$ while there is no dimerization when $\delta=0$.
The latter system can be interpreted as a precursor of a metal since in the thermodynamic limit this system becomes metallic.
It is convenient to express the Hamiltonian in the site basis.
Upon diagonalization we thus obtain the eigenfunctions which, when inserted in Eqs.~(\ref{Eqn:TPS_OBC}) and (\ref{Eqn:TPS_PBC}), yield
the OBC and PBC localisation tensors of the dimerized chain.

\begin{figure}[t]
\begin{center}
\includegraphics[width=\columnwidth]{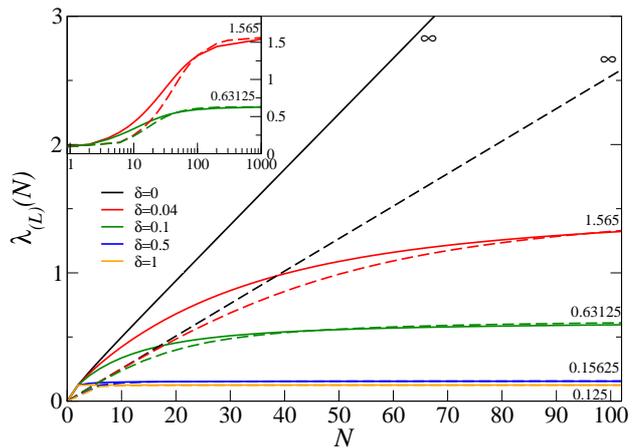}
\end{center}
\caption{The OBC localisation tensor $\lambda(N)$ (solid lines) and the PBC localisation tensor $\lambda_L(N)$ (dashed lines) 
as a function of the number of electrons $N$ for various values of the dimerization parameter $\delta$ in a tight-binding model (see Eq.~(\ref{Eqn:HTB})).
The numbers next to each curve are the values of the PBC localisation tensor in the thermodynamic limit.
They were obtained from Eq.~(\ref{Eqn:Huckel}) with $d=1$.
Inset: $\lambda_{(L)}(N)$ for $\delta=0.04$ and $\delta=0.1$ for large $N$.}
\label{Fig:lambda}
\end{figure}

In Fig.~\ref{Fig:lambda} we report $\lambda(N)$, as defined in Eq.~(\ref{Eqn:TPS_OBC}) for OBC, 
as well as $\lambda_{L}(N)$, as defined in Eq.~(\ref{Eqn:TPS_PBC}) for PBC, 
as a function of the number of electrons for various values of $\delta$.
The length of the unit cell has been set to unity and we have set $L=N$ such that the density $N/L=1$.
First of all, we see that the PBC localization tensor is well-defined for finite $N$ also for the "metallic" chain ($\delta=0$).
Second, as expected, the OBC and PBC localisation tensors, $\lambda(N)$ and $\lambda_L(N)$, respectively, differ for finite $N$, 
since they describe different systems.
Instead, in the thermodynamic limit both localisation tensors describe the same system and indeed we obtain the same values in that limit.

The advantage of the PBC localisation tensor is that the thermodynamic limit can be obtained without extrapolating results for finite $N$ and $L$.
Instead, at least for single-particle Hamiltonians, one can obtain an expression involving information of a single unit cell.

In the case of the dimerized chain one can derive an analytical expression for the 
localisation tensor in the thermodynamic limit. The unit cell of length $d$ contains two sites 
separated by $d/2$ with one electron per site. In the site basis the Hamiltonian $\hat{H}_{\kappa}$, 
which corresponds to the periodic part of the wave function, is then given by ($d=1$):
\begin{align}
H_{\kappa,11} &= H_{\kappa,22} = 0 \\
H_{\kappa,12} &= H_{\kappa,21}^* = -(1+\delta)e^{\frac{- \pi \imun  }{\kappa n}}  +(1-\delta)e^{\frac{ \pi \imun  }{\kappa n}}
\end{align}
where $\kappa=0,1,2, \ldots, n-1$ is an integer and $n$ is the number of cells.
The eigenvectors of the matrix $H_\kappa$ are the periodic part of the Bloch functions $\phi_{ik}$.
There is one occupied valence state $\phi_{vk}$ and one unoccupied conduction state $\phi_{ck}$.
They are given by
\begin{equation}
\phi_{ik}(x)  = \frac{1}{\sqrt{n}} e^{\imun  k x} u_{ik}(x) \;\; (i=v,c)
\end{equation}
where $x$ is the coordinate along the chain, $k=2 \pi \kappa /L$ is the wave vector and
$u_{ik}(x)$ is a periodic function that is normalized over a single cell.
To obtain the localisation tensor we use a similar strategy to that used for the system 
of non-interacting electrons, \emph{i.e.}, 
we insert a complete set of states in Eq.~(\ref{Eqn:TPS_PBC}) and use the Slater-Condon rules.
This yields the following expression~\cite{monari2008}:
\begin{equation}
\lambda_L(N) = 
\frac{L^2}{4\pi^2}
\sum_{\kappa=0}^{n-1} \langle\phi_{vk}| e^{-i kx}|\phi_{ck}\rangle
\langle\phi_{ck}|  e^{i kx} |\phi_{vk}\rangle.
\end{equation}
In the thermodynamic limit the variable $\kappa$ becomes continuous and the summation over $\kappa$ 
can be replaced with an integral over $k$. 
%
We finally obtain
\begin{equation}
\lambda_{\infty} \equiv \lim_{N,L\rightarrow\infty}\lambda_L(N) = \frac{1+\delta^2 } {16 |\delta|},
\label{Eqn:Huckel}
\end{equation}
As expected, for a periodic system, $\lambda_{\infty}$ is an even function of $\delta$.
We note that $\lambda_\infty$ goes to infinity as $1/|\delta|$ when $\delta$ tends to zero, \emph{i.e.}, when we go from an insulator
($\delta\neq 0$) to a metal ($\delta=0$). 
We reported the limiting values obtained with Eq.~(\ref{Eqn:Huckel}) in Fig.~\ref{Fig:lambda}.

In conclusion, we have presented a simple one-body position operator that is compatible with periodic boundary conditions.
We have shown that this operator meets several important fundamental constraints, \emph{e.g.}, 
it is translationally invariant, it reduces to the common position operator $\hat{x}=x$ in the appropriate limit, and
the distance expressed in terms of the operator is gauge independent.
Moreover, we have demonstrated its usefulness when applied to the localisation tensor.
In particular, we have shown that it yields the correct expression in the thermodynamic limit 
and correctly distinguishes between finite precursors of insulators and metals.
Finally, we note that this work opens the road for the calculation of various other properties of interest 
for which the corresponding operators involve the position operator.
In particular, the proposed definition of position could be used in order to reconcile Coulomb's law with periodic boundary conditions. 
Preliminary investigations in this direction, both on the quantum ({\em ab initio} treatment of the electron gas) and classical level (Madelung sums) are in progress.

This work was partly supported by the French
``Centre National de la Recherche Scientifique'' (CNRS, also under the PICS action 4263).
It has received fundings from the European Union's Horizon 2020 research and innovation programme under the Marie Sk{\l}odowska-Curie grant agreement n\textsuperscript{o}642294.
This work was also supported by the ``Programme Investissements d'Avenir'' under the
program ANR-11-IDEX-0002-02, reference ANR-10-LABX-0037-NEXT.
The calculations of this work have been partly performed by using the resources of the HPC center CALMIP under the grant 2016-p1048.
Finally, one of us (EAV) acknowledges the support of the 
``Theoretical Chemistry and Computational Modeling(TCCM) Erasmus-Plus Master program.

\end{document}